\documentclass{iau} 
\usepackage{graphicx}

\title[Statistical Properties of Superflares on Solar-type Stars] %% give here short title %%
{Statistical properties of superflares on solar-type stars based on the Kepler 1-min cadence data}

\author[Maehara, H., Shibayama, T., Notsu, Y., et al.]   %% give here short author list %%
{Hiroyuki Maehara$^{1,*}$, 
 Takuya Shibayama$^2$,
Yuta Notsu$^{3}$,
Shota Notsu$^{3}$,
Satoshi Honda$^{4}$,
Daisaku Nogami$^{3}$,
\and Kazunari Shibata$^5$}

\affiliation{
$^*$email: {\tt h.maehara@oao.nao.ac.jp} \\[\affilskip]
$^1$Okayama Astrophysical Observatory, National Astronomical Observatory of Japan, 3037-5 Honjo, Kamogata, Asakuchi, Okayama, Japan, 719-0232 \\
$^2$Institute for Space-Earth Environmental Research, Nagoya University, Furo-cho, Chikusa-ku, Nagoya, Aichi, Japan, 464-8601\\
$^3$Department of Astronomy, Kyoto University, Kitashirakawa-Oiwake-cho, Sakyo-ku, Kyoto, Japan, 606-8502 \\
$^4$Center for Astronomy, University of Hyogo, 407-2, Nishigaichi, Sayo-cho, Sayo, Hyogo, Japan, 679-5313 \\
$^5$Kwasan and Hida Observatories, Kyoto University, Yamashina-ku, Kyoto, Japan, 607-8471
}

\pubyear{2015}
\volume{320}  %% insert here IAU Symposium No.
\jname{Solar and Stellar Flares and Their Effects on Planets}
\editors{A.G. Kosovichev, S.L. Hawley \& P. Heinzel, eds.}
\begin{document}

\maketitle

\begin{abstract}
We searched for superflares on solar-type stars using the Kepler 
short-cadence (1-min sampling) data in order to detect superflares 
with short duration. 
We found 187 superflares on 23 solar-type stars whose bolometric
energy ranges from the order of $10^{32}$ erg to $10^{36}$ erg.
%Some superflares show multiple peaks with the peak
%separation of the order of 100-1000 seconds which is comparable 
%to the periods of quasi-periodic pulsations in solar and stellar flares.
Using these new data combined with the results from the data with 
30-min sampling, we found the occurrence frequency ($dN/dE$) of 
superflares as a function of flare energy ($E$) shows the power-law 
distribution ($dN/dE \propto E ^{-\alpha}$)
with $\alpha=1.5$ for $10^{33}<E<10^{36}$ erg.
%The average occurrence rate of superflares with the energy of 1033 erg on
%Sun-like stars (early G-dwarfs with rotation periods comparable to that of the Sun) which is equivalent to X100 solar
%flares is estimated to be about once in 500-600 years.
The upper limit of energy released by superflares is basically
comparable to a fraction of the magnetic energy stored near starspots
which is estimated from the amplitude of brightness variations.
We also found that the duration of superflares ($\tau$) increases with 
the flare energy ($E$) as $\tau \propto E^{0.39\pm 0.03}$. This can be
explained if we assume the time-scale of flares is determined by the 
Alfv$\acute{\rm e}$n time.
\end{abstract}

\firstsection % if your document starts with a section,
\section{Introduction}
 Solar-flares are energetic eruption in the solar atmosphere caused by
the magnetic reconnection (e.g. \cite[Shibata \& Magara 2011]{Shibata2011}).
The occurrence frequency of solar-flares ($dN/dE$) as a function of 
flare energy ($E$)
can be fitted by a simple power-law function ($dN/dE \propto E ^{-\alpha}$)
with $\alpha=1.5$ - $1.9$
in the flare energy range between $10^{24}$ erg and $10^{32}$ erg
(e.g. \cite[Crosby et al. 1993, Shimizu 1995, Aschwanden et al. 2000]{Aschwanden2000, Shimizu1995, Crosby1993}).
The bolometric energy released by an X10 class solar flare
is estimated to be the order of $10^{32}$ erg (\cite[Emslie et al. 2012]{Emslie2012})
and the occurrence frequency of such flare is about once in 10 years.
In the case of other stars, including solar-type stars,
more energetic flares called ``superflares" have been observed (e.g. \cite[Landini et al. 1986, Schaefer 1989, Schaefer et al. 2000]{Landini1986,Schaefer1989, Schaefer2000}).
Recently, many superflares on solar-type stars (G-type main sequence stars) 
have been discovered by the Kepler space telescope (e.g. \cite[Maehara et al. 2012, Shibayama et al. 2013]{Maehara2012, Shibayama2013}).
The bolometric energy released by the superflare ranges from $10^{33}$ to 
$10^{36}$ erg which is $10$-$10^4$ times larger than that of the largest solar flares ($\sim 10^{32}$ erg).
Majority of solar-type stars exhibiting superflares show
quasi-periodic brightness variations with amplitude range from
$\sim 0.1$\% to $8$\%.
These light variations are thought to be caused by the rotation of the star 
with starspots (e.g. \cite[Notsu et al. 2013]{Notsu2013b}).
Although most of solar-type stars with superflares are rapidly rotating stars, 
the temperature and rotation velocity of some superflare stars are close 
to those of the Sun (\cite[Nogami et al. 2014]{Nogami2014}).

Previous studies of superflares on solar-type stars using the Kepler
long-cadence data mainly focus on 
superflares with extremely large energy ($E > 10^{34}$ erg) and 
long duration ($> 1$ hour).
Here we report the results from the data with high time-resolution 
($\sim 1$ min) and discuss the statistical properties (e.g. occurrence 
frequency, flare duration) of superflares on solar-type stars.
More details are described in \cite{Maehara2015}.

\section{Data}
We searched for flares 
from the short-cadence data ($\sim 1$ min interval) observed with the 
Kepler space telescope between 2009 April (quarter 0: Q0) and 
2013 May (Q17) (\cite[Koch et al. 2010, Gilliland et al. 2010]{Koch2010, Gilliland2010}).
We selected solar-type (G-type main sequence) stars from the data set
by using the surface temperature of the star ($T_{\rm eff}$) and 
the surface gravity ($\log g$) taken from \cite{Huber2014}
instead of those from \cite{Brown2011} (initial Kepler Input Catalog).
In previous papers (\cite[Maehara et al. 2012, Shibayama et al. 2013]{Maehara2012, Shibayama2013}), 
we used stellar parameters taken from \cite{Brown2011}
and the selection criteria of $5100 \rm{K} < T_{\rm eff} < 6000$ K and 
$\log g > 4.0 $.
However the temperatures in \cite{Brown2011} are
systematically lower by $\sim 200$ K than those in \cite{Huber2014}.
In order to reduce the difference in statistical properties of superflares
caused by the systematic difference in stellar temperature, 
here we used the selection criteria of $5300 \rm{K} < T_{\rm eff} < 6300 \rm{K}$
and $4.0 < \log g < 4.8$.
The total number of solar-type stars observed with short-cadence mode is
1547.

We used the Presearch Data Conditioned (PDC) light curve (\cite[Stumpe et al. 2012, Smith et al. 2012]{Stumpe2012, Smith2012})
for the detection of flares.
The detail procedures of the flare detection are described in \cite{Maehara2015}.
Bolometric energy released by each flare was estimated 
from the flare amplitude, flare duration, and stellar luminosity 
with the same manner as \cite{Shibayama2013}.
The stellar luminosity were calculated from the effective temperature 
and the stellar radius taken from \cite{Huber2014}.
We estimated periods of long-term light variations from light curves 
of each star by using the discrete Fourier transform (DFT) method.
For period analysis, we used the long-cadence data (time resolution of 30 min)
obtained from 2009 September (Q2) to 2013 April (Q16).
These periods are basically consistent with those in \cite{McQuillan2014}.

\section{Results and discussion}
\begin{figure}[b]
% \vspace*{-2.0 cm}
\begin{center}
 \includegraphics[width=\linewidth]{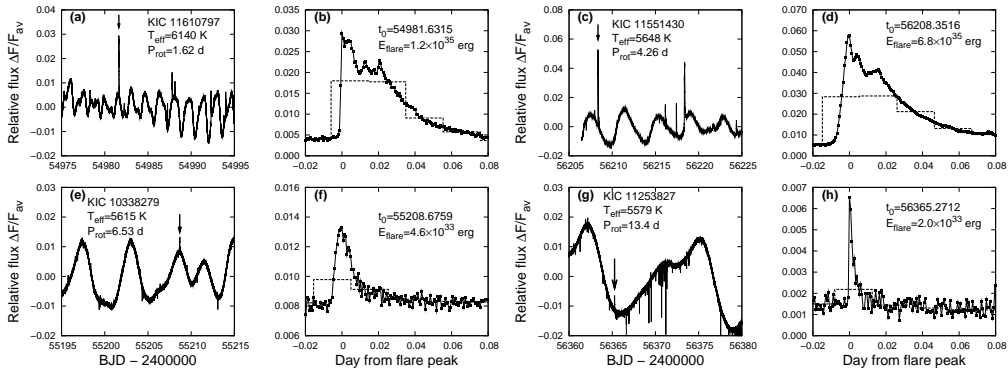} 
% \vspace*{-1.0 cm}
 \caption{
(a), (c), (e), (g): Long-term light variations of the G-dwarf KIC 11610797 (a), 
KIC 11551430 (c), KIC 10338279 (e), and KIC 11253827 (g).
The vertical axis
means the relative difference between observed brightness of the star
and the average brightnes during the observation period.
The horizontal axis means the times of the observations in 
Barycentric Julian Date.
(b), (d), (f), (h): Enlarged light curve of a superflare on KIC 11610797 (b),
KIC 11551430 (d), KIC 10338279 (f), and KIC 11253827 (h) (indicated by the down arrow in panel (a), (c), (e), and (g)).
Filled-squares with solid-lines and dashed-lines represent
the light curves from  short- and long-cadence data respectively.
}
   \label{lc}
\end{center}
\end{figure}

We detected 187 flares on 23 solar-type stars 
from the data of 1547
solar-type stars.
Figure \ref{lc} shows typical light curves of superflares 
detected from short-cadence data.
The amplitude of flares normalized by the average brightness of the star
range from $1.3\times 10^{-3}$ to $8.5\times 10^{-2}$
and the bolometric energies of flares range from $2\times 10^{32}$ to $8\times 10^{35}$ erg.

\subsection{Occurrence frequency distribution}
\begin{figure}[b]
% \vspace*{-2.0 cm}
\begin{center}
 \includegraphics[width=\linewidth]{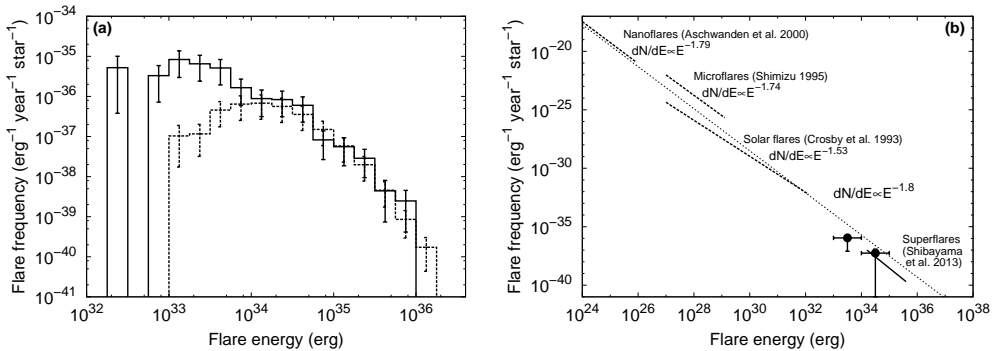} 
% \vspace*{-1.0 cm}
 \caption{
(a): Bold solid and dashed histograms represent the occurrence frequency 
of superflares on all solar-type stars from short- (this work) and 
long-cadence data (\cite[Shibayama et al. 2013]{Shibayama2013}) as a
function of total energy of superflares respectively.
The vertical axis indicates the number of superflares per star, per year, and per unit energy.
(b): Comparison between occurrence frequency superflares on Sun-like stars and those of solar flares. 
Filled-circles indicates the occurrence frequency of superflares on 
Sun-like stars (solar-type stars with $P_{\rm rot}>10$ days and 
$5800<T_{\rm eff}<6300$K) derived from short-cadence data.
Bold-solid line represents the power-law frequency distribution of 
superflares on Sun-like  stars taken from \cite{Shibayama2013}.
Dashed lines indicate the power-law frequency distribution of solar 
flares observed in hard X-ray (\cite[Crosby et al. 1993]{Crosby1993}), 
soft X-ray microflares (\cite[Shimizu 1995]{Shimizu1995}), 
and EUV nanoflaers (\cite[Aschwanden et al. 2000]{Aschwanden2000}).
Occurrence frequency distributions of superflares on Sun-like stars 
and solar flares are roughly on the same power-law line 
with an index of $-1.8$ (thin-solid line; \cite[Shibata et al. 2013]{Shibata2013}) for the wide energy 
range between $10^{24}$ erg and $10^{35}$ erg.
}
   \label{frequency}
\end{center}
\end{figure}

Figure \ref{frequency} (a) represents the occurrence frequency of 
superflares ($dN/dE$) as a function of the bolometric energy of superflares ($E$).
Solid histogram in figure \ref{frequency} (a) represents the 
frequency distribution of superflares on all solar-type stars derived 
from short-cadence data and dashed histogram represents that from 
long-cadence data (\cite[Shibayama et al. 2013]{Shibayama2013}).
Since the period distribution of the stars observed in short-cadence 
mode is biased to the shorter-period end, we estimate the corrected
occurrence frequency for short-cadence data taking into account 
the bias in the period distribution of the observed samples.
Both frequency distributions from short- and long-cadence data
 are almost the same for the flare energy between $10^{34}$ erg 
and $10^{36}$ erg and can be fitted by a power-law function 
($dN/dE \propto E ^{-\alpha}$).
Using the combined data set from both short- and long-cadence data,
 the power-law index $\alpha$ is $1.5 \pm 0.1$
for the flare energy of $4\times 10^{33}$ - $1\times 10^{36}$ erg.

According to \cite{Shibata2013}, the frequency distribution of 
superflares on Sun-like stars (early-G dwarfs with $P_{\rm rot} \geq 10$ day)
and those of solar flares are roughly on the same power-law line.
Figure \ref{frequency} (b) represents the comparison between the 
frequency distribution of superflares on Sun-like stars 
($5800 \leq T_{\rm eff} < 6300$ and $P_{\rm rot} \geq 10$ day) 
derived from both  short- (filled circles) and long-cadence 
(solid line) data and those of solar flares (dashed lines).
The thin dotted line indicates the power-law function ($dN/dE \propto E ^{-\alpha}$) with an index $\alpha=1.8$ taken from \cite{Shibata2013}.

\subsection{Flare energy and area of starspots}
Majority of solar-type stars exhibiting superflares show
quasi-periodic light variations with the amplitude of the order of 1 \%.
These variations are thought to be caused
by the rotation of star with spotted surface and
the amplitude of light variations corresponds to 
the area of starspots.
Figure \ref{spotsize} shows the scatter plot of the energy of solar flares
and superflares
as a function of the area of sunspot/starspot group ($A_{\rm spot}$).
The area of starspots on superflare stars were estimated from the 
amplitude of light variations.% ($\Delta F_{\rm rot}$) from the 
%following equation:
%\begin{equation}
%A_{\rm spot} = \Delta F_{\rm rot} A_{\rm star} \left\{1-\left(\frac{T_{\rm spot}}{T_{\rm star}}\right)^{4}\right\}^{-1}, 
%\end{equation}
% where $A_{\rm star}$ is the apparent area of the star, $T_{\rm spot}$ and $T_{\rm star}$ are the temperature of starspot and photosphere of the star (\cite[Notsu et al. 2013]{Notsu2013b}).
%$A_{\rm star}$ and $T_{\rm star}$ for each superflare star were taken from \cite{Huber2014} and we assumed $T_{\rm spot} = 4000$K (e.g. \cite[Berdyugina 2005]{Berdyugina2005}).
%The majority of energetic superflares occur on the stars with large starspots.
Since the flares are a result of a sudden conversion of magnetic energy 
to thermal and kinetic energy,
the bolometric energy released by the flare should  be a fraction of the 
magnetic energy ($E_{\rm mag}$) (\cite[(Shibata et al. 2013, Notsu et al. 2013)]{Shibata2013, Notsu2013b}).
%The stored magnetic energy can be roughly estimated by
%\begin{equation}
%E_{\rm mag} \sim \frac{B^2 L^3}{8\pi} ,
%\end{equation}
%where $B$ and $L$ correspond to the magnetic field strength of the starspots 
%and the scale length of the starspot group respectively.
%Since the total area of starspot group can be written as,
%$A_{\rm spot}\sim L^{2}$, 
The energy released by a flare can be written as the following form:
\begin{equation}
E_{\rm flare} \sim f E_{\rm mag} \sim \frac{fB^2 L^3}{8\pi} \sim \frac{f B^2 A_{\rm spot}^{3/2}}{8\pi} ,
\label{e_flare}
\end{equation}
where $f$, $B$ and $L$ correspond to the fraction of magnetic energy released by the flare, the magnetic field strength of the starspots 
and the scale length of the starspot group respectively.
The equation (\ref{e_flare}) suggests that the flare energy is roughly proportional to the area of
the starspot group to the power of 3/2.
According to \cite{Aschwanden2014}), $f \sim 0.1$.
The typical sunspot area for generating X10-class flares 
($E_{\rm flare} \sim 10^{32}$ erg)
observed in 1989-1997  was $3\times  10^{-4}$ of the 
half area of the solar surface (\cite[Sammis et al. 2000]{Sammis2000}),
and the typical magnetic field strength of sunspot is the 
order of $1000$ G (e.g. \cite[Antia et al. 2003]{Antia2003}). Therefore
equation (\ref{e_flare}) can be written as
\begin{equation}
E_{\rm flare} \sim 7\times 10^{32} {\rm (erg)} \left(\frac{f}{0.1}\right) \left(
\frac{B}{1000{\rm G}}\right)^2 \left[\frac{A_{\rm spot}/2\pi R_{\odot}^2}{0.001}
\right]^{3/2} .
\label{e_flare2}
\end{equation}
Solid and dashed lines in the figure \ref{spotsize} represent 
equation \ref{e_flare2} for $f=0.1$, $B=1000$ G and $f=0.1$, $3000$ G 
respectively. 
Majority of superflares detected from short-cadence data (filled-squares)
and almost all solar flares (small dots) are below these analytic lines. 
This suggests that the upper limit of the energy released by the 
flare is basically comparable to
the stored magnetic energy estimated from the area of starspots.
However there are some superflares above the analytic relation.
These superflares would occur on the stars with low-inclination angle 
or stars with starspots around polar region. In the case of such stars,
the light variation caused by the rotation become small. Hence
the the actual area of starspots should be larger than that estimated 
from the amplitude of brightness variation.
According to \cite{Notsu2015b}, some superflare stars which show 
superflares with energy
larger than that expected from the amplitude of light variation were
confirmed to have low-inclination angles.
Moreover polar spots are often observed on rapidly-rotating 
late-type stars (e.g. \cite[Strassmeier 2009]{Strassmeier2009}).

\subsection{Correlation between flare duration and flare energy}
\begin{figure}[b]
\begin{minipage}[t]{0.48\linewidth}
% \vspace*{-2.0 cm}
\begin{center}
 \includegraphics[width=\linewidth]{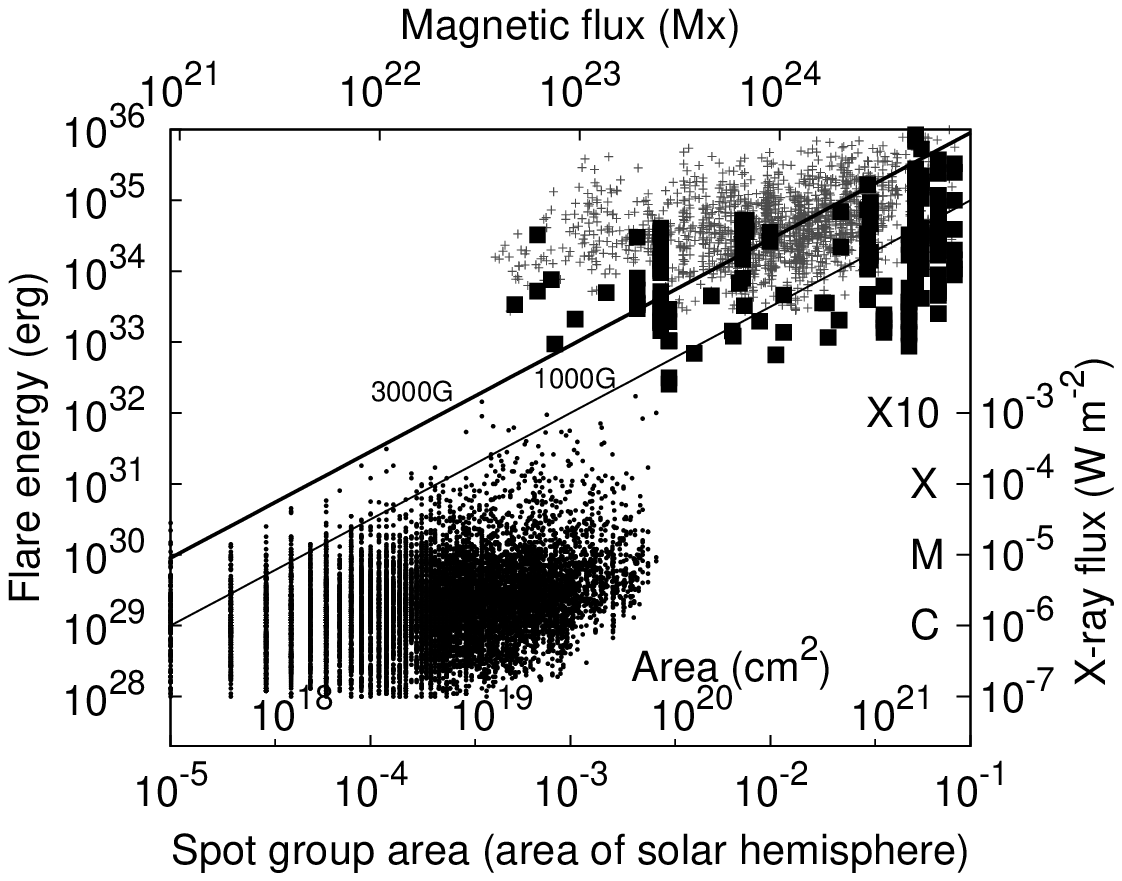} 
% \vspace*{-1.0 cm}
 \caption{
Scatter plot of flare energy as a function of spot area.
The lower and upper horizontal axis indicate the area of starspot group in the u
nit of the area of solar hemisphere and the magnetic flux for $B=3000$ G.
The vertical axis represents the bolometric energy released by each flare.
Filled-squares and  small-crosses indicate 
superflares on solar-type stars from short-(this work) and long-cadence data (\cite[Shibayama et al. 2013]{Shibayama2013}) respectively.
mall filled-circles 
represent solar flares 
(Ishii et al., private communication, based on the data
retrieved from the website of the National Geophysical 
Data Center of the National Oceanic and Atmospheric Administration 
(NOAA/NGDC), Solar-Terrestrial Physics Division at
http://www.ngdc.noaa.gov/stp/).
}
   \label{spotsize}
\end{center}
\end{minipage}
\begin{minipage}[t]{0.02\linewidth}
\end{minipage}
\begin{minipage}[t]{0.48\linewidth}
% \vspace*{-2.0 cm}
\begin{center}
 \includegraphics[width=\linewidth]{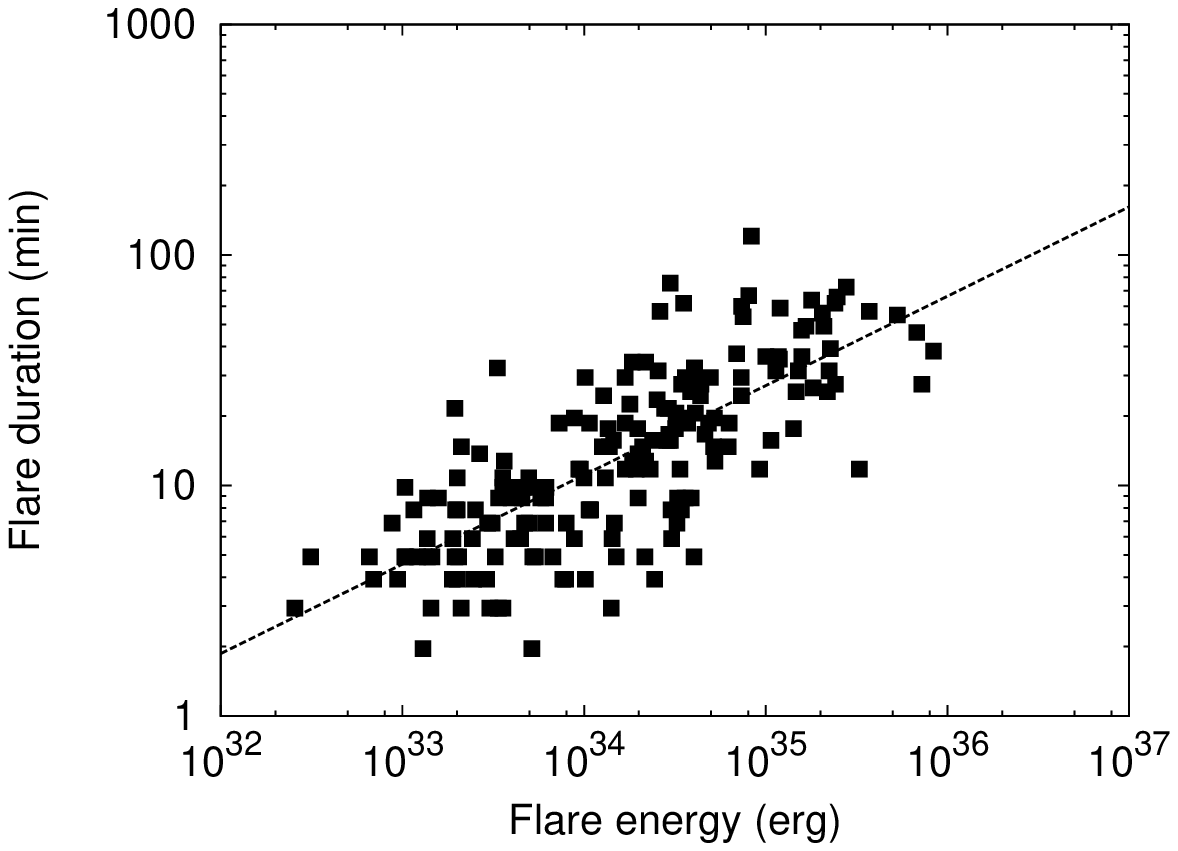} 
% \vspace*{-1.0 cm}
 \caption{
Scatter plot of the duration of superflares as a function of the 
bolometric energy.
Filled-squares indicate superflares on solar-type stars
from short-cadence data.
We used $e$-folding decay time as the flare duration.
Dotted line indicates the linear regression for the data of superflares from short-cadence data. The power-law slope of the line is $0.39 \pm 0.03$
}
   \label{duration}
\end{center}
\end{minipage}
\end{figure}

Figure \ref{duration} represents the duration ($e$-folding decay time) 
of superflares as a function of the energy of flares.
A linear fit for the superflares from short-cadence data in the 
$\log$-$\log$ plot yields
\begin{equation}
\tau _{\rm flare} \propto E^{0.39\pm 0.03}_{\rm flare}
\end{equation}
where $\tau _{\rm flare}$ and $E_{\rm flare}$ indicate the 
duration and energy of superflares.
Similar correlation between the flare duration and energy was observed 
in solar flares (e.g. \cite[Veronig et al. 2002, Christe et al. 2008]{Veronig2002, Christe2008}).
The flare energy is 
related to the magnetic energy stored near the starspots as shown in
equation \ref{e_flare}.
On the other hand, the duration of impulsive phase of flares
is comparable to the reconnection time ($\tau _{\rm rec}$) 
which can be written as
\begin{equation}
\tau _{\rm flare} \sim \tau _{\rm rec} \sim \tau _{\rm A}/M_{\rm A} \sim L/v_{\rm A}/M_{\rm A},
\label{flare_duration}
\end{equation}
where $\tau _{\rm A} = L/v_{\rm A}$ is the Alfv$\acute{\rm e}$n time, 
$v_{\rm A}$ is the Alfv$\acute{\rm e}$n velocity, and $M_{\rm A}$ is
the non-dimensional reconnection rate ($M_{\rm A} \sim 0.1$-$0.01$ for the fast
reconnection; \cite[Shibata \& Magara 2011]{Shibata2011}).
If we assume $B$ and $v_{\rm A}$ are not so different among solar-type stars,
the duration of flares can be written as
\begin{equation}
\tau _{\rm flare} \propto E_{\rm flare} ^{1/3}.
\end{equation}
This suggests that the power-law slope for the 
correlation between the flare duration 
and flare energy is about $1/3$ which is comparable to the 
observed value of $0.39\pm 0.03$.

\section{Conclusions}
We found 187 superflares on 23 solar-type stars from the Kepler
1-min cadence data. This new data set of superflares on solar-type
stars suggests that
the power-law frequency distribution of superflares (e.g. 
\cite[Maehara et al. 2012, Shibayama et al. 2013]{Maehara2012, Shibayama2013}) 
continues to the flare energy of $10^{33}$ erg.
We also found the flare duration increases with the flare energy as 
$\tau_{\rm flare} \propto E_{\rm flare}^{0.39\pm 0.03}$. This suggests
that the time-scale of flares is determined by the Alfv$\acute{\rm e}$n time.

\end{document}